\documentclass[showpacs,preprintnumbers,amsmath,amssymb,superscriptaddress,floatfix,nofootinbib]{revtex4}
\usepackage{epsfig}
\long\def\symbolfootnote[#1]#2{\begingroup%
\def\thefootnote{\fnsymbol{footnote}}\footnote[#1]{#2}\endgroup}
\newcommand{\dslash}[1]{#1\!\!\!/}

  \def\CL{{\cal L}}
\def\CM{{\cal M}}

\def\npa#1{Nucl.\ Phys.\ A\ {\bf #1}}

\def\plb#1{Phys.\ Lett.\ B\ {\bf #1}}

\def\etal{{\em et al.}}

\def\be{\begin{equation}}
\def\ee{\end{equation}}
\def\Be{\begin{eqnarray}}
\def\Ee{\end{eqnarray}}
\def\ba{\begin{array}}
\def\ea{\end{array}}

\begin{document}
\title{Resonance model study on $K^+ N \rightarrow K N \eta$ near threshold}

\author{Bo-Chao Liu}
\email{liubc@xjtu.edu.cn}\affiliation{Department of Applied Physics,
Xi'an Jiaotong University, Xi'an, Shaanxi 710049, China}

\begin{abstract}

By using resonance model, we investigate $K^+ N \rightarrow K N \eta$ reactions with the assumption
 that these reactions are
dominated by the excitation of $N^*(1535)$ near threshold. It is found
that the hyperon and $\rho$ exchange diagrams give the  most
important contributions to these reactions. Thus these reactions may be a good place to
study the coupling of $N^*(1535)$ with $K\Lambda$, $K\Sigma$ and $N\rho$ channels. We demonstrate that the angular distributions of final particles provide useful information about the different mechanisms of the $N^*(1535)$ excitations, which could be useful for future experimental analysis on these reactions.

\end{abstract}
\maketitle

\section{Introduction}
The negative parity nucleon resonance $N^*(1535)$ is particularly
interesting in light hadron physics because of its peculiar
properties. It is the chiral partner ($J^P=\frac{1}{2}^-$) of the
nucleon, and has strong decay channels for both $\pi N$ and $\eta
N$. Although it is ranked as a four-star state in PDG\cite{pdg2010},
the nature and property of $N^*(1535)$ are still not well
understood. Besides the conventional constituent quark model
interpretation, it has also been argued that $N^*(1535)$ is a
quasi-bound($K\Sigma -K\Lambda$)-state\cite{Weise} and has large
effective couplings to $K\Lambda$ and $K\Sigma$\cite{Oset}. To check
these model predictions, experimental information on the coupling of
$N^*(1535)$ with $KY$(kaon-hyperon) states should be necessary.
Unfortunately current experimental knowledge on these kaon-hyperon
couplings is still poor, partly because of lack of data on
experimental side and partly due to the complication of various
interfering t-channel exchange contributions\cite{Mosel} in $\pi N$
and $\gamma N$ scatterings.

In recent years, the decay of $J/\Psi$ is also utilized to study the
properties of nucleon resonances\cite{NsatBes}. Because the isospin
of $J/\Psi$ is zero, its decay offers a natural isospin filter which
makes it a unique place to study the properties of nucleon
resonances. In the reaction $J/\Psi \rightarrow p K^- \bar\Lambda$,
it is found that there is an enhancement in $K\Lambda$ invariant
mass spectrum near threshold\cite{bes3}. If this enhancement is
caused by $N^*(1535)$, it will imply that $N^*(1535)$ has a
large coupling to $K\Lambda$ and will have important
implications on the property and nature of
$N^*(1535)$\cite{liu1,liu2}. Based on a similar picture, it is
also argued that $N^*(1535)$ probably has large coupling to
$N\phi$\cite{xie,shi,cao}. Obviously, some further studies on
the coupling of $N^*(1535)$ with $KY$ states will be helpful to
understand the nature of $N^*(1535)$ and relevant reaction
mechanisms.

Besides the $KY$ couplings, the coupling of $N^*(1535)$ with vector
meson and nucleon is also not well determined, which causes the debate that
whether $\pi$\cite{bati,naka,shyam07}
    or $\rho$\cite{geda,san,faldt,naka02,vetter}meson exchange diagram dominates
   $\eta$
production in nucleon-nucleon collisions. The
   differences between these two kinds of models are generally related to the uncertainties of the coupling
    constant $g_{N^*(1535)N\rho}$. Even though the
   uncertainties of  this coupling constant are examined in detail in Ref.~\cite{xie2}, it is still interesting
   and important
   to constrain the value of this coupling constant in some other channels.

   With the problems mentioned above, it is natural to ask whether there are some other channels which are suitable for studying the
   properties of $N^*(1535)$. In this work, by using resonance model we study the reactions
   $K^+ p \rightarrow K^+ p \eta$, $K^+ n \rightarrow K^0 p \eta$ and $K^+ n \rightarrow K^+ n \eta$ with the assumption that the
   excitation of $N^*(1535)$ dominates these reactions near threshold.
 Some other contributions to this channel are mainly from the excitation of the $K^*$ resonances and other nucleon resonances besides
  $N^*(1535)$. Because we are only interested
 in the energy range near threshold, it is reasonable to expect that only the states which have S-wave coupling to $K\eta$ or $N\eta$ channel can give significant contributions.
 For $K\eta$ channel, the $K^*$ state in relevant energy range that has S-wave coupling with $K\eta$ is $K^*_0(1430)$, which is about 400 MeV above $K\eta$ threshold and should have minor
 effects near threshold. Furthermore, there are also some indirect evidences  from the Dalitz plots\cite{keta1,keta2} show
 that $K^*$s do not give significant contribution near threshold. For the subthreshold contribution from $K^*(892)$,
  we note that it has p-wave coupling to $K\eta$ and its mass is about 150MeV below threshold.
   In view of its relatively small width, i. e. 50 MeV, we expect that the contribution from $K^*(892)$
    should also have minor effects near threshold.   Meanwhile, according to PDG\cite{pdg2010}, we find that
near $\eta N$ threshold the $S_{11}$ states $N^*(1535)$ and $N^*(1650)$ have
significant decay branch ratios to both $N\rho$ and $N\eta$ channels
and may give sizable
 contributions to these reactions. With the parameters and formulas
 offered in Ref.~\cite{caoxu}, we calculate the contribution from
 $N^*(1650)$ and find its contribution is very small compared to the
 contribution from $N^*(1535)$ because of its larger mass
 and relatively weak coupling with $N\eta$ channel. The dominance of
 $N^*(1535)$ in $N\eta$ channel near threshold is also well identified in relevant
 experimental studies of $J/\Psi \rightarrow p \bar p \eta$\cite{BES}
 and $ p p\rightarrow pp\eta$\cite{cosy} reactions. Based on the considerations given above, we ignore the contribution
 from $K^*$s and other nucleon resonances in present work.
And, due to no clear evidence of the existence of
pentaquark,
 we also ignore the s-channel pentaquark contributions.

In next section, we will give the formalism and ingredients in our
calculation, and then numerical results and some discussions are
given in Sec. III. A short summary is given in the last section.

\section{Theoretical Formalism}
 \begin{figure}[htbp] \vspace{0.cm}
\begin{center}
\includegraphics[scale=0.5]{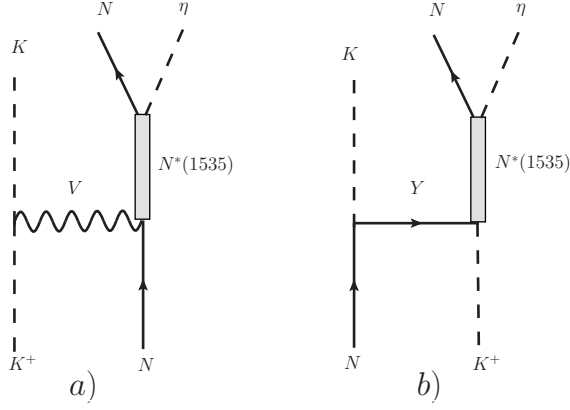}
 \caption{Feynman diagrams for the excitation of $N^*(1535)$ through a) vector meson exchange and b) hyperon
  exchange in $K^+ N \rightarrow K N \eta$ reactions. For $K^+ p \rightarrow K^+ p
  \eta$ :
 V=$\rho^0$, $\phi$ or $\omega$ and Y=$\Lambda$ or $\Sigma^0$; for $K^+ n \rightarrow K^0 p \eta$ : V=$\rho^{\pm}$ and Y=$\Lambda$ or $\Sigma^0$;
 for $K^+ n \rightarrow K^+ n \eta$ : V=$\rho^0$, $\phi$ or $\omega$ and Y=$\Sigma^-$. }
 \label{feyn_fig}
\end{center}
\end{figure}
In this work, we study the reactions $K^+ p \rightarrow K^+ p \eta$,
$K^+ n \rightarrow K^0 p \eta$ and $K^+ n \rightarrow K^+ n \eta$ in
an effective Lagrangian approach. We assume that these reactions are
dominated by the intermediate
 excitation of the $N^*(1535)$ near threshold and then $N^*(1535)$ decays to $N\eta$. The basic Feynman diagrams for the $K^+ N\rightarrow K N \eta$  are depicted
 in Fig.~\ref{feyn_fig}.

We use the commonly used interaction Lagrangians for $\rho
KK$, $\omega KK$ and $\phi KK$ couplings\cite{VPP}, \Be
\CL_{\rho K \bar K}&=&iG_V{[\bar K \vec{\tau}(\partial_\mu K)-(\partial_\mu \bar K)\vec{\tau}K]\cdot\vec{\rho}^\mu}\\
\CL_{\omega K \bar K}&=&iG_V{[\bar K (\partial_\mu K)-(\partial_\mu \bar K)K]\omega^\mu}\\
\CL_{\phi K \bar K}&=&-\sqrt{2}iG_V{[\bar K (\partial_\mu
K)-(\partial_\mu \bar K)K]}\phi^\mu
 \Ee At each vertex a relevant off-shell form factor is used. In our
computation, we take the same form factors as that widely
used\cite{VPPFF} \be
F^{KK}_{V}=\frac{\Lambda^2_V-m_V^2}{\Lambda^2_V-q_V^2} \ee where
$\Lambda_V$, $m_V$ and $q_V$ are the cutoff parameter, mass and
four-momentum for the exchanged meson(V) respectively. We adopt the
coupling constant $G_V$ and $\Lambda_V$ as $G_V=3.02$ and
$\Lambda_V=2$ GeV in the calculations\cite{VPPFF}.

To calculate the Feynman diagrams in Fig.~\ref{feyn_fig}, we still need to know the
interaction Lagrangian involving $N^*(1535)$ resonance. In
Ref.~\cite{zouprc03}, a Lorentz covariant orbital-spin (L-S) scheme
for $N^*NM$ couplings has been given in detail. With this scheme, we
can easily write the effective $N^*(1535)N\eta$, $N^*(1535)N\rho$,
$N^*(1535)N\omega$ and $N^*(1535)N\phi$ vertex functions, \Be
V_{N^*(1535)N\eta}&=&ig_{N^*(1535)N\eta}\bar u_Nu_{N^*(1535)} +h.c.,\\
V_{N^*(1535)N\rho}&=&ig_{N^*(1535)N\rho}\bar u_N\gamma_5 \left (\gamma_\mu-\frac{q_\mu\gamma^\nu q_\nu}{q^2}\right ) \varepsilon^\mu(p_\rho)u_{N^*(1535)}+h.c.,\\
V_{N^*(1535)N\omega}&=&ig_{N^*(1535)N\omega}\bar u_N\gamma_5 \left (\gamma_\mu-\frac{q_\mu\gamma^\nu q_\nu}{q^2}\right ) \varepsilon^\mu(p_\omega)u_{N^*(1535)}+h.c.,\\
V_{N^*(1535)N\phi}&=&ig_{N^*(1535)N\phi}\bar u_N\gamma_5 \left
(\gamma_\mu-\frac{q_\mu\gamma^\nu q_\nu}{q^2}\right )
\varepsilon^\mu(p_\phi)u_{N^*(1535)}+h.c.. \Ee Here $u_N$ and
$u_{N^*}$ are the spin wave functions for the nucleon and
$N^*(1535)$ resonance; $\varepsilon^\mu(p_\rho)$,
$\varepsilon^\mu(p_\omega)$ and $\varepsilon^\mu(p_\phi)$ are the
polarization vectors of the $\rho$, $\omega$ and $\phi$ mesons,
respectively. It is worth noting that because the spin of vector
meson is 1, both S-wave and D-wave L-S couplings are possible for
the $N^*(1535)N\rho$, $N^*(1535)N\omega$ and $N^*(1535)N\phi$
interactions. It was found that the S-wave coupling has significant
contribution to the $N^*(1535)$ decaying to $N\rho$ compared with
the D-wave \cite{vrana,pdg2010}. In our calculations we consider
only the S-wave $N^*(1535)$ resonance coupling to $N\rho$ and
neglect the D-wave coupling. We also neglect the D-wave $N^*(1535)$
resonance couplings to $N\omega$ and $N\phi$ for simplicity as it
was done in Ref.~\cite{xie,shi}. The monopole form factors for
$N^*(1535) N$-meson vertices are used, \be
F^{N^*}_{NM}=\frac{\Lambda^{*2}-m_V^2}{\Lambda^{*2}-q_V^2} \ee where
$m_V$ and $q_V$ are the mass and four momentum of the exchanging
vector mesons and we adopt $\Lambda^* = 1.3$~GeV \cite{xie} in our
work. For the coupling constant $g_{N^*(1535)N\rho}$, we take
$g^{2}_{N^*(1535)N\rho}/4\pi =0.1$\cite{xie} in our calculation
which is determined by the partial decay width
$\Gamma_{N^*(1535)\rightarrow N\rho \rightarrow N\pi\pi}$. It is
shown in Ref.~\cite{xie2} that this value is also consistent with
the prediction of the radiation decay of $N^*(1535)$ within vector
meson dominance model. For the coupling constant $g_{N^*(1535)
N\eta}$, we use the value $g^2_{N^*(1535)
N\eta}/4\pi=0.28$\cite{xie}, which is obtained from the partial
decay width of $N^*(1535)$ to $N\eta$. The coupling constant
$g_{N^*(1535)N\omega}$ is still not well constrained by experimental
data. In the literatures, the ratio of $g_{N^*(1535)N\rho}$ to
$g_{N^*(1535)N\omega}$ varies from 1.77 to
2.6\cite{N1535Nomega1,N1535Nomega2,N1535Nomega3}.  In this work, we
adopt the value of ratio as 2, which gives
$g^2_{N^*(1535)N\omega}/4\pi=0.25$. Another coupling constant
$g_{N^*(1535)N\phi}$ is also not well known. However, in
Ref.~\cite{xie} it is shown that if assuming a large coupling of
$N^*(1535)$ with $N\phi$, both $\pi^- p\rightarrow n \phi$ and
$pp\rightarrow pp \phi$ data can be well described. So in this work,
we adopt $g^2_{N^*(1535)N\phi}/4\pi=0.13$ as suggested in
Ref.~\cite{xie}. And concrete calculations show that, even with this
large coupling constant, $\phi$ exchange diagram only plays a minor
role in these reactions.

The other class of Feynman diagram considered in this work is Fig.~\ref{feyn_fig}b. The
effective Lagrangians describing the couplings of $N^*(1535)$ to $KY$ and $N$ to $KY$
are taken from Ref.~\cite{liu1,liu2} \Be
\CL_{N^*(1535)K\Lambda}&=&-ig_{N^*(1535)K\Lambda}\bar \Psi_{N^*(1535)}\Psi_\Lambda \Phi_K +h.c.,\\
\CL_{N^*(1535)K\Sigma}&=&-ig_{N^*(1535)K\Sigma}\bar \Psi_{N^*(1535)}\vec{\tau}\cdot\vec\Psi_{\Sigma}\Phi_K +h.c.,\\
\CL_{NK\Lambda}&=&ig_{NK\Lambda}\bar \Psi_N\gamma_5 \Psi_{\Lambda}\Phi_K +h.c.\\
\CL_{NK\Sigma}&=&ig_{NK\Sigma}\bar \Psi_N\gamma_5
\Psi_{\Sigma}\Phi_K +h.c. \Ee For the value of coupling constants
$g_{NK\Lambda}$ and $g_{NK\Sigma}$, one popular choice is to use
SU(3) predictions. It has been shown that $pp\rightarrow
pKY$\cite{gknl1,gknl2} and $Kp$ scattering\cite{gknl3} can be
understood in terms of $g_{K\Lambda p}$ and $g_{K\Sigma p}$ values
which are in good agreement with the SU(3) predictions. Also from a
Regge analysis of the high energy $\gamma p \rightarrow KY$
data\cite{gknl4}, it seems that these coupling constants are still in
agreement with SU(3) predictions. So we adopt the SU(3) predicted
values, i.e. $g_{NK\Lambda}^2/4\pi=14.06$ and
$g_{NK\Sigma}^2/4\pi=1.21$, in our calculations. For the coupling
constants $g_{N^*K\Lambda}$ and $g_{N^*K\Sigma}$, one option is to
determine them from SU(3) predictions, because it was shown in
Ref.~\cite{N1535Lambda1670} that the SU(3) relations may hold for
$N^*(1535)$. Within this option, one can follow the logic and
results given in Ref.~\cite{su3}. With the parameters given in that
work, i.e. $\alpha=-0.28$ and $|A_8|=5.2$, we get
$g_{N^*(1535)K\Lambda}^2/4\pi=0.14$ and
$g_{N^*(1535)K\Sigma}^2/4\pi=5.24$(Option I). The other option is to
follow the results given in Ref.~\cite{Oset}, where the ratio
between $|g_{N^*(1535)K\Lambda}|$, $|g_{N^*(1535)K\Sigma}|$ and
$|g_{N^*(1535)N\eta}|$ can be obtained as $0.92 : 1.5 : 1.84$. With
the value of $g_{N^*(1535)N\eta}$ given above, then we get
$g_{N^*(1535)K\Lambda}^2/4\pi=0.069$ and
$g_{N^*(1535)K\Sigma}^2/4\pi=0.19$(Option II). By comparing these
two options, we find that for the coupling constant
$g_{N^*(1535)K\Lambda}$ these two options give some similar
predictions. While, for the coupling constant
$g_{N^*(1535)K\Sigma}$, the predictions from these two options are
very different. The SU(3) prediction for $g_{N^*(1535)K\Sigma}$ is
about 5.3 times larger than that obtained from Ref.~\cite{Oset}.
With this uncertainty in mind, we adopt the Option II in the
following calculations, because with a very large
$g_{N^*(1535)K\Sigma}$ it may cause problems in consistently
describing some other relevant processes, such as $\gamma p\to
K\Sigma$ or $\pi^- p \to K\Sigma$ reactions, where $N^*(1535)$ also
contributes. The final conclusion on the value of
$g_{N^*(1535)K\Sigma}$ should be made with a thorough analysis of
all relevant channels. The form factors for the vertices $NKY$ and
$N^*KY$ are taken from Ref.~\cite{uchannel} \be
F_{KY}=\frac{\Lambda^4_u}{\Lambda^4_u+(q^2-m^2)^2}. \label{uff} \ee For
the cut off parameter of vertex $KN\Lambda$, it is known that to
control the Born amplitudes of reaction $\gamma p\rightarrow K^+
\Lambda$ in a reasonable range the introduction of mechanism that
reduces the Born strength is necessary\cite{uchannel}. One possible
way is to introduce a rather small $\Lambda_u$,
 and it is shown that the experimental data can be described fairly well with $\Lambda_u=0.4$ GeV\cite{lam1,lam2}.
 However, with such a small $\Lambda_u$,
the form factors play a predominant role in the reaction dynamics
and may cause serious questions about the validity of theoretical
framework. Also, using such a small value of $\Lambda_u$ one
cannot give consistent descriptions of the reaction $e p \rightarrow e K^+ \Lambda$
as well\cite{ep2eklambda}. So we adopt $\Lambda_u=1.5$ GeV
 for vertex $KN\Lambda$ in our work as suggested in Ref.~\cite{uchannel}. To reduce the number of free parameters, we use the same
 cutoff parameter for the vertex $KN\Sigma$ as well. For the vertices
$N^*(1535)K\Lambda$ and $N^*(1535)K\Sigma$ in u-channel, we use the
same form factor as that defined in Eq.(\ref{uff}). However, the cut
off parameter ($\Lambda^*_u$ ) for these vertices are not well
determined in the literatures. In this work, we adopt
$\Lambda^*_u=1.3$ GeV for these vertices, and the uncertainties due
to this parameter will be discussed below. For easy comparison with
other works, all the coupling constants and cut-off parameters
adopted in our work are collected in Tab.\ref{cc}.
\begin{table}
\caption{\label{table}Coupling constants and cut-off parameters adopted in present work.}
\begin{center}
\begin{tabular}{|ccc|ccc|}
\hline
\hline Vertex & g  & $\Lambda$[GeV]&Vertex&$g^2/4\pi$& $\Lambda$[GeV]\\
\hline
$\rho KK$ & $G_V=3.02$ & 2.0  &$N^*(1535)N\rho$ &  0.1 & 1.3    \\
$\omega KK$ & $G_V$ &  2.0 & $N^*(1535)N\omega$  &  0.25 & 1.3   \\
$\phi KK$ & $\sqrt{2}G_V$ & 2.0 &$N^*(1535)N\phi$  & 0.13 & 1.3     \\
$N K\Lambda$ & $-$13.29 & 1.5 &$N^*(1535) K\Lambda$ & 0.069 & 1.3     \\
$N K\Sigma$ & 3.9 & 1.5  & $N^*(1535) K\Sigma$  & 0.19 & 1.3    \\
            &       &   &  $N^*(1535)N\eta$  & 0.28   &   \\
\hline
\hline
\end{tabular}
\end{center} \label{cc}
\end{table}

The $N^*(1535)$ propagator is written in a Breit-Wigner
form~\cite{liang}:
\begin{equation}
G_{N^*}(q)=\frac{i(\dslash{q}
+M_{N^*})}{q^2-M^2_{N^*}+iM_{N^*}\Gamma_{N^*}(q^2)}\,,
\end{equation}
where $\Gamma_{N^*}(q^2)$ is the energy-dependent total width and q is the four momentum of $N^*(1535)$.
Keeping only the dominant $\pi N$ and $\eta N$ decay
channels~\cite{pdg2010}, this can be decomposed as
\begin{equation}
\Gamma_{N^*} (q^2) = a_{\pi N}\, \rho_{\pi N}(q^2) + b_{\eta N} \,
\rho_{\eta N}(q^2),
\end{equation}
where $a_{\pi N} = 0.12$~GeV/$c^2$, $b_{\eta N} = 0.32$~GeV/$c^2$,
and the two-body phase space factors, $\rho_{\pi(\eta)N}(q^2)$, are
\begin{equation}
\rho(q^2)= \left.2
p^{\,\text{cm}}(q^2)\,\Theta(q^2-q_{\text{thr}}^2)\right/\!\!\sqrt{q^2}\,,
\label{psf}
\end{equation}
and $q_{\text{thr}}$ is the threshold value for the decay
channel.

The propagators of vector meson and hyperon are also needed in the
calculations and can be written in the form \Be G^{\mu
\nu}_{V}(q_{V})&=&- i (\frac{g^{\mu
\nu}-q^{\mu}_{V}q^{\nu}_{V}/q^2_{V}}{q^2_{V}-m^2_{V}}) \\
G_Y(q_{Y})&=&i \frac{\dslash{q}_Y+M_{Y}}{q^2_{Y}-m^2_{Y}} . \Ee
where $q_{V}$ and $q_Y$ are the 4-momentum of the exchanged vector
meson and hyperon($Y=\Lambda$ or $\Sigma$) respectively.

After having established the effective Lagrangians, coupling
constants and form of the propagators, the invariant scattering
amplitudes can be written by following the standard Feynman rules.
The calculations of the differential and total cross sections are
then straightforward,
\begin{eqnarray}
d\sigma=\frac{1}{16}\frac{m_N^2}{\sqrt{(p_K\cdot
p_N)^2-m_N^2m_K^2}}\frac{1}{(2 \pi)^5} \sum_{s_i,s_f} |{\cal
M}_{fi}|^2\prod^3_{a=1}\frac{ d^{3} p_{a}}{E_{a}} \delta^4 (P_i -
P_f), \label{eqcs}
\end{eqnarray}
where $\CM_{fi}$ represents the total amplitude, $P_i$ and $P_f$
represent the sum of all the momenta in the initial and final
states, respectively, and $p_a$ denotes the momenta of the three
particles in the final state.

\begin{figure}[htbp] \vspace{0.cm}
\begin{center}
\includegraphics[scale=0.3]{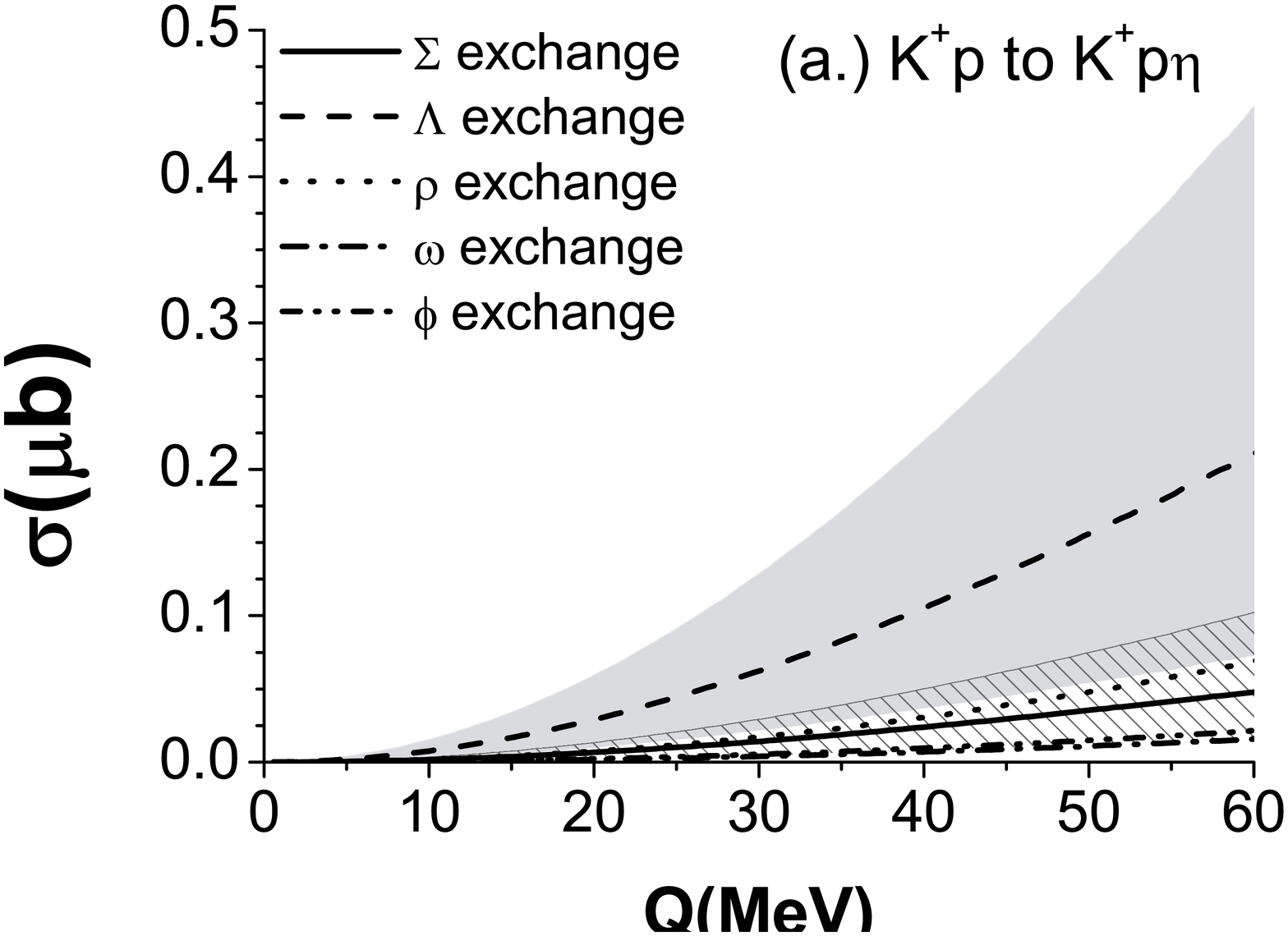}
\includegraphics[scale=0.3]{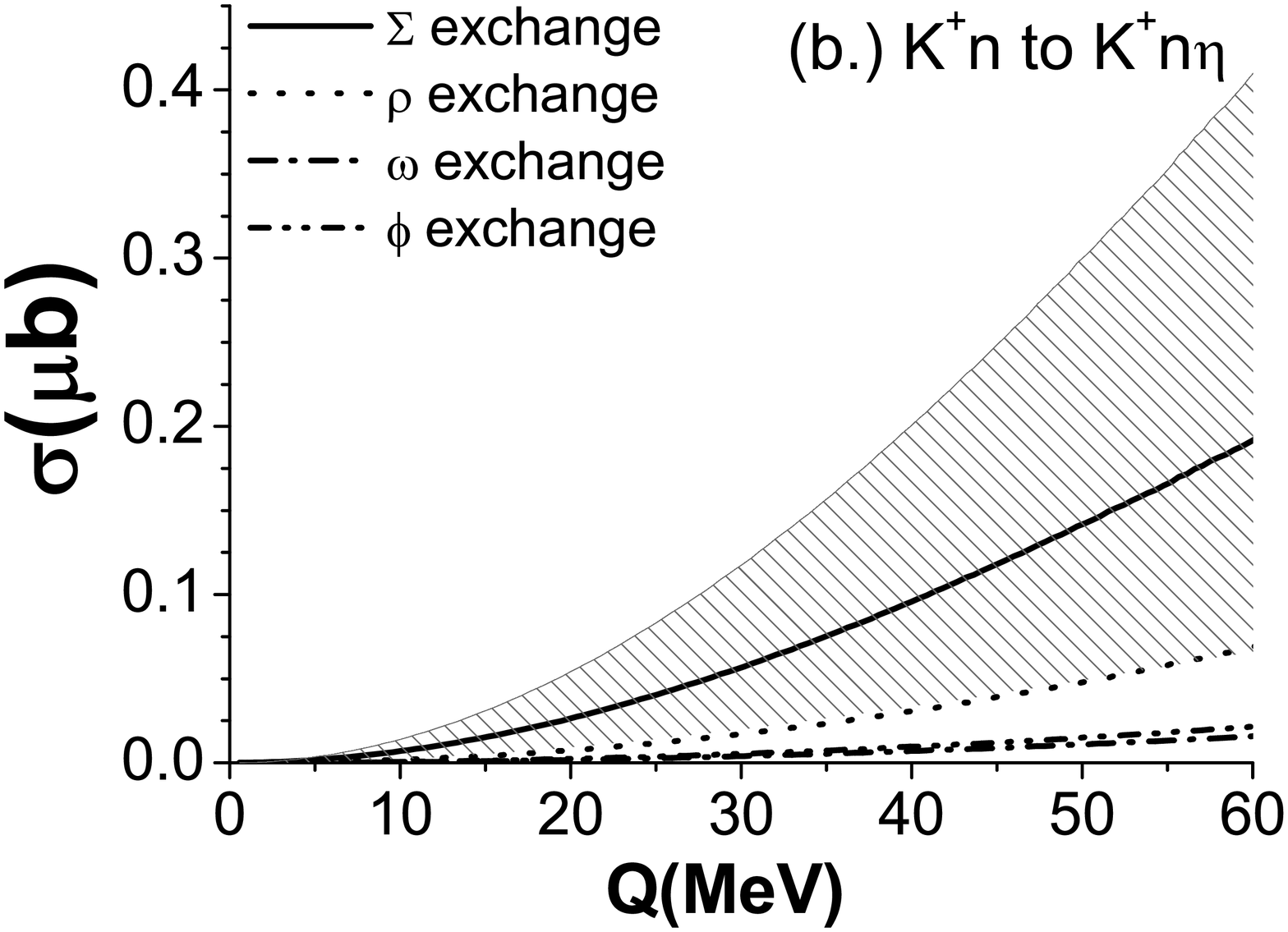}
\includegraphics[scale=0.3]{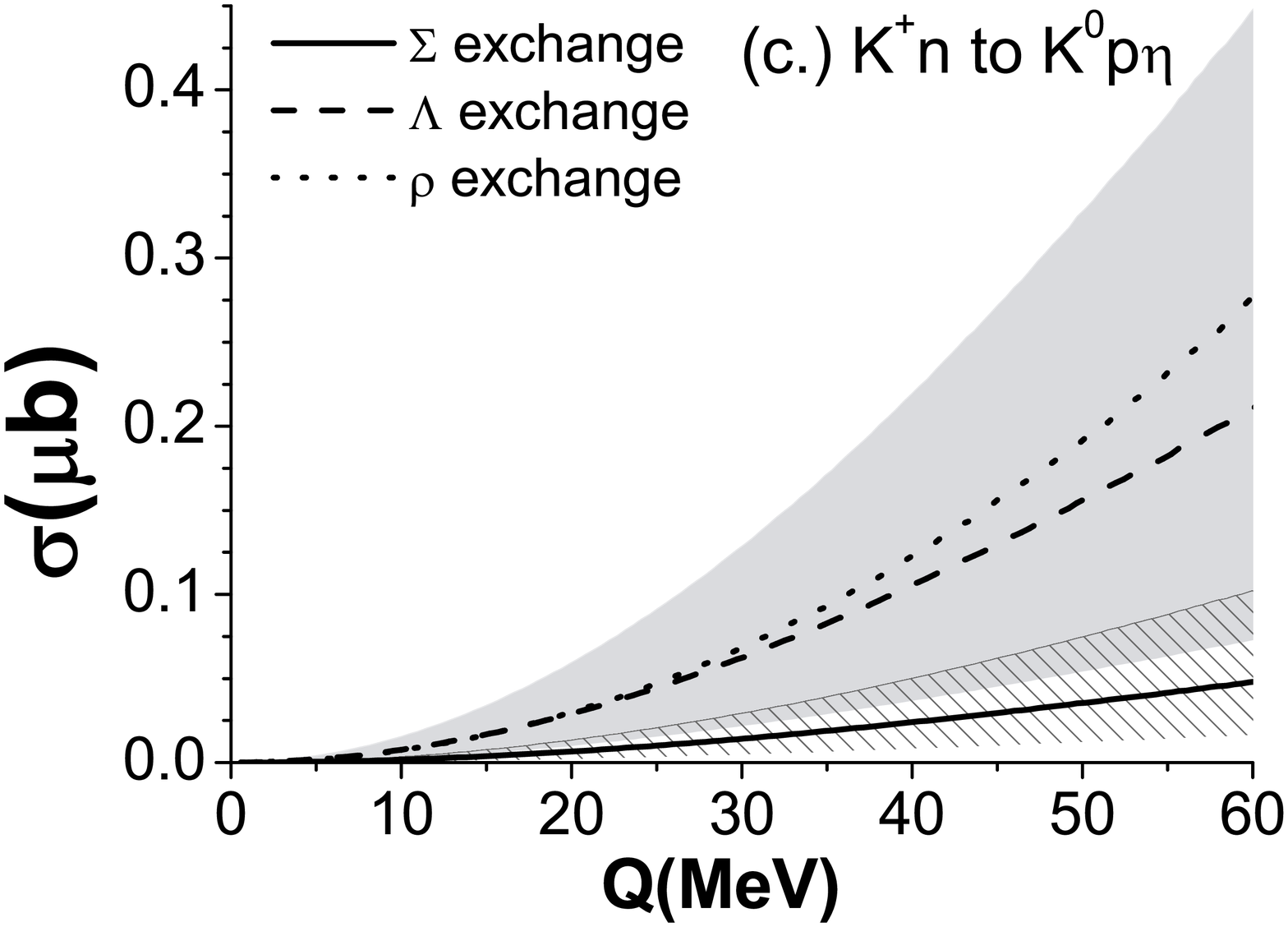}
 \caption{ The total cross section obtained by considering individual diagrams for reactions (a) $K^+ p \rightarrow K^+ p
\eta$, (b) $K^+ n \rightarrow K^+ n \eta$ and (c) $K^+ n \rightarrow
K^0 p \eta$, where the gray and shadowed areas denote the
uncertainties due to the cut-off parameters on $N^*(1535)K\Lambda$
and $N^*(1535)K\Sigma$ respectively.}
 \label{xsection}
\end{center}
\end{figure}

 \begin{figure}[htbp] \vspace{0.cm}
\begin{center}
\includegraphics[scale=0.8]{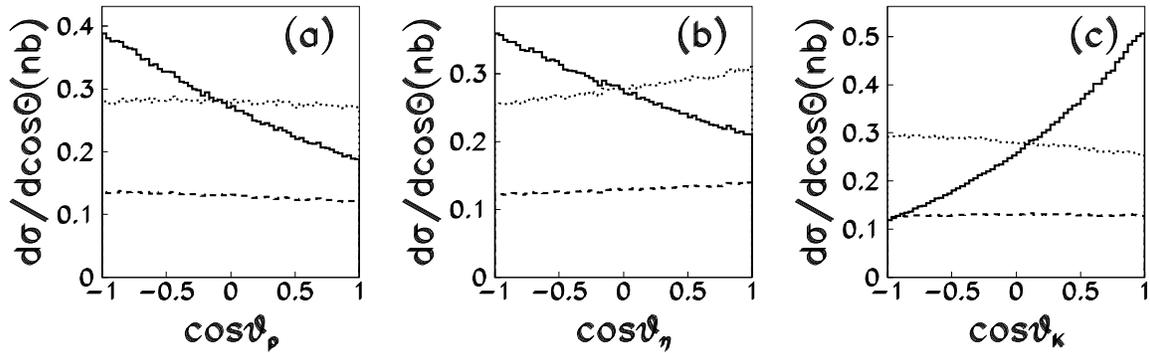}
 \caption{Angular distribution of final proton (a), $\eta$ (b) and $K^0$ (c) of the reaction $K^+ n \rightarrow K^0 p \eta$, where $\theta$ denotes the angle of the outgoing particles relative to the incident $K^+$ beam direction in c.m. frame.   The solid, dashed and dotted lines represent the contribution from $\rho$, $\Sigma$ and $\Lambda$ exchange amplitudes respectively.}
 \label{ang_dis}
\end{center}
\end{figure}
\section{RESULTS AND DISCUSSIONS}

With the formalism and ingredients given above, the total cross
sections versus excess energy Q for the $K^+ p \rightarrow K^+ p
\eta$, $K^+ n \rightarrow K^+ n \eta$ and $K^+ n \rightarrow K^0 p
\eta$ are calculated by using a Monte Carlo multi-particle phase
space integration program.  In Fig.~\ref{xsection}, we show the
results of cross sections obtained by considering vector meson
exchange and hyperon exchange diagrams.

From Fig.~\ref{xsection}, it can be found that the $\Lambda$, $\Sigma$
and $\rho$ exchanges give the most important contributions to these
reactions.  The $\phi$ exchange contribution only plays a minor role, although we adopt a large value for
$g_{N^*(1535)N\phi}$. The strength of $\omega$ exchange is a little
smaller than $\phi$ exchange within our model. In $K^+ p \rightarrow
K^+ p \eta$, the $\Lambda$ exchange dominates this
reaction near threshold. The contribution from $\Sigma$ exchange is
much smaller than $\Lambda$ exchange, which is mainly due to the
large difference between the values of $g_{KN\Lambda}$ and
$g_{KN\Sigma}$. While for the reaction $K^+ n \rightarrow K^+ n
\eta$, $\Sigma$ exchange plays the most important role. This is
partly because $\Lambda$ exchange is forbidden in this reaction and
partly because $\Sigma$ exchange is enhanced in this channel because of
the isospin Clebsch-Gordan coefficients appearing in the vertex
functions. In the reaction $K^+ n \rightarrow K^0 p \eta$, $\omega$
and $\phi$ exchanges are forbidden and $\rho$ exchange becomes more
important compared to other channels. The $\rho$ exchange gives
equally important contribution as $\Lambda$ exchange. It is also because of
the isospin Clebsch-Gordan coefficients appearing in the vertices
that make $\rho$ exchange much more favored in this reaction.

 To check the dependence of the results on the cutoff parameter $\Lambda^*_u$ adopted for $N^*(1535)KY$ vertex,
  we also perform the calculations with $\Lambda^*_u=1.0$ GeV
 and $\Lambda^*_u=2.0$ GeV respectively.  With a smaller cutoff value, i.e. 1.0 GeV, the
strength of the $\Lambda$ and $\Sigma$ exchange amplitudes is suppressed and their
contributions to cross section are
 reduced by a factor of
 3.
 However, with
$\Lambda^*_u=2.0$ GeV, the contributions from $\Lambda$ and $\Sigma$
exchanges are enhanced by a factor of 2. The uncertainties due to
this parameter, which are obtained by varying the $\Lambda^*_u$ from 1.0 GeV to 2.0 GeV, are shown in
Fig.~\ref{xsection} by the gray and shadowed area for $\Lambda$
exchange and $\Sigma$ exchange respectively. The error bands show
that the value of this cut off parameter is important for
determining the magnitudes of amplitudes. Unfortunately, because cut
off parameter is introduced phenomenologically, it only can be
determined by fitting to experimental data. Without experimental
data near threshold, this parameter cannot be well determined in
present model. However, the above calculations may offer us some
estimation about the uncertainties of present model.
 \begin{figure}[htbp] \vspace{0.cm}
\begin{center}
\includegraphics[scale=0.8]{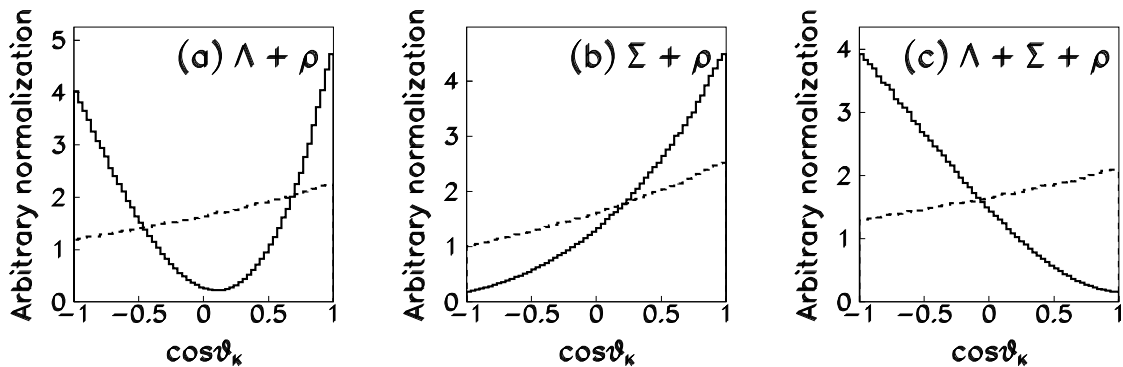}
 \caption{Illustration of the interference effects on the angular distribution of final K meson in c.m. frame in $K^+ n \rightarrow K^0 p \eta$ reaction by considering the interference among $\rho$, $\Sigma$ and $\Lambda$ exchange amplitudes.   The solid and dashed line represent the results corresponding to destructive and constructive interference respectively. $\theta$ is defined in the same way as in Fig.~\ref{ang_dis}.}
 \label{interference}
\end{center}
\end{figure}

With the uncertainties mentioned above,  it still can be expected
from the results shown in Fig.~\ref{xsection} that the $\Lambda$,
$\Sigma$ and $\rho$ exchanges play the most important roles in the
reactions $K^+ N \rightarrow K N \eta$ near threshold. Thus these
reactions may constitute a good basis for investigating the
couplings of $N^*(1535)$ with $KY$ and $N\rho$ channels. Due to the
charge conservation law, $\Lambda$ exchange is forbidden in $K^+ n
\rightarrow K^+ n \eta$. And because $\Sigma$ exchange dominates
this reaction, this reaction may be a good place to extract the
coupling constant $g_{N^*(1535)K\Sigma}$. Similarly, because
$\Lambda$ and $\rho$ exchanges give the most important contributions
to the reactions $K^+ p \rightarrow K^+ p \eta$ and $K^+ n
\rightarrow K^0 p \eta$, these reactions are suitable to study the
coupling constants $g_{N^*(1535)K\Lambda}$ and $g_{N^*(1535)N\rho}$.

 To distinguish the contributions between hyperon
and $\rho$ exchanges, one possible way is to utilize the angular
distribution of final particles in center of mass frame(c.m. frame). To
illustrate this possibility, we show the angular distribution of
final particles in the reaction $K^+ n \rightarrow K^0 p \eta$ in
center of mass system at Q=15 MeV in Fig.~\ref{ang_dis} by considering
$\Lambda$, $\Sigma$ and $\rho$ exchanges respectively, where $\rho$
and $\Lambda$ exchanges have similar strength and $\Sigma$ exchange
only plays a minor role. Note that here we choose the energy Q=15 MeV
just as an example, and the angular distributions do not change
significantly near threshold. As can be seen from Fig.~\ref{ang_dis},
the angular distribution from $\rho$ exchange amplitude and hyperon
exchange amplitude are distinct from each other.  The pattern of
angular distributions shown in Fig.~\ref{ang_dis} can be understood
in the following way. If we ignore the decay of $N^*(1535)$, the
$\rho$ and hyperon exchange diagrams are corresponding to t-channel
and u-channel diagrams respectively. So one may expect that the
angular distribution of final K meson should have a forward peak for
the $\rho$ exchange amplitude and have a backward peak for hyperon
exchange amplitude respectively. And this is what we get in
Fig.~\ref{ang_dis}.

In order to investigate the interference effects, we need to fix the relative
phase among individual amplitudes which in principle should be done
by fitting to the data within an effective Lagrangian approach. To get an estimation
 of the interference effects, in this work
we assume that the relative phase between $\Lambda$ exchange
amplitude and $\Sigma$ exchange amplitude is fixed by the SU(3)
symmetry, i.e. we adopt the SU(3) predicted sign for the relevant
coupling constants. The relative phase between $\rho$ exchange
amplitude and $\Lambda$ exchange amplitude is taken to be either
$+1$ or $-1$ corresponding to the constructive and destructive
interference respectively. In this way, we can fix the relative
phases among individual amplitudes and the corresponding results for
the angular distributions are shown in Fig.~\ref{interference}, where we present the results by considering the
coherent sum of the $\rho$ and $\Lambda$
exchanges(Fig.~\ref{interference}a),  $\rho$ and $\Sigma$
exchanges(Fig.~\ref{interference}b), and the full
mechanisms(Fig.~\ref{interference}c) respectively. It needs to be
noted that significant interference effects among individual
amplitudes are also found in the calculation of total cross
sections. However, in order to show the interference effects on the
shape of the angular distribution more clearly,  we normalize
individual results to the same quantity. As can be seen from
Fig.~\ref{interference}, the interference between individual
mechanisms may alter the angular distribution considerably compared
to the distribution from the individual mechanisms without interference effects in
Fig.~\ref{ang_dis}. This shows clearly that the interference
effects may have important influence on the physical observables.

Based on the above discussions, it can be expected that the experimental
 data of angular distributions may present very different pattern as compared
 to the angular distributions shown in Fig.~\ref{ang_dis} where
 interference effects are not taken into account. This will make it difficult to extract the relevant
couplings from the experimental data directly. Here we want to note that the strength and relative roles of $\Sigma$ and $\rho$ exchanges change
 in different reactions. This means that if
 the angular distributions are sensitive to the relative phase and magnitude
 of individual amplitudes as shown in Fig.~\ref{interference}, the angular distributions
 of final particles would vary significantly in different reactions.
Because the three reactions considered in this work are related by
isospin symmetry, the strength of individual mechanism in different reactions is related by isospin relations. Thus
a combined analysis of all these three
reactions can put strong constraints on the magnitude and relative
phase of individual amplitudes, which will help us understand the
coupling of $N^*(1535)$ with various channels better. And the
specific features of angular distributions due to individual mechanisms given in present work
could be helpful for analyzing the reaction mechanisms when
experimental data are available.
\section{Summary}
 In this work, we study the reactions $K^+ N
\rightarrow K N \eta$ near threshold within an effective Lagrangian
approach. Based on the assumption that this reaction is dominated by
the excitation of $N^*(1535)$ resonance, we find that the $\Lambda$,
$\Sigma$ and $\rho$ exchange diagrams give the most important
contributions to these reactions near threshold. Thus the reactions
under study may constitute a good basis to study the coupling of
$N^*(1535)$ with $N\rho$, $K\Lambda$ and $K\Sigma$ channels. It is
also found that interference effects among individual mechanisms are
important and may alter the angular distributions significantly. A
combined analysis on all the three reactions can help us better
understand the relative roles of individual mechanisms, and the
results of this work should be useful for analyzing and entangling
the different mechanisms when the experimental data are available in
the future.

\begin{acknowledgements}
We acknowledge Zhen Ouyang and Ju-Jun Xie for their careful reading the manuscript and
useful suggestions. This work is supported by the National Natural Science
Foundation of China under Grant No. 10905046.
\end{acknowledgements}

\end{document}